\documentclass[epj,twocolumn]{webofc}
\usepackage[colorlinks=true]{hyperref}
\usepackage{color}
\usepackage[varg]{txfonts}   
\usepackage{amssymb}	
\usepackage{bm}         
\hypersetup{citecolor=blue, linkcolor=blue, urlcolor=blue}
\usepackage{graphicx}

\newcommand{\pt}{{p_{\mathrm t}}}

\wocname{epj}
\woctitle{Seismology of the Sun and the Distant Stars 2016}
\begin{document}
\title{Calibrating damping rates with LEGACY\thanks{data provided by Mikkel N. Lund (e-mail: mikkelnl@phys.au.dk)} linewidths}
%

\author{\firstname{Günter} \lastname{Houdek}\inst{1}\fnsep\thanks{\email{hg@phys.au.dk}}
}

\institute{Stellar Astrophysics Centre, Aarhus University, DK-8000 Aarhus C, Denmark 
}

\abstract{%
Linear damping rates of radial oscillation modes in selected $Kepler$ stars are
estimated with the help of a nonadiabatic stability analysis. The convective fluxes
are obtained from a nonlocal, time-dependent convection model. The mixing-length
parameter is calibrated to the surface-convection-zone depth of a stellar model obtained
from fitting adiabatic frequencies to the LEGACY$^\star$ observations, and two of the three
nonlocal convection parameters are calibrated to the corresponding LEGACY$^\star$ 
linewidth measurements. The atmospheric structure in the 1D stability analysis 
adopts a temperature-optical-depth relation derived from 3D hydrodynamical 
simulations. Results from 3D simulations are also used to calibrate the 
turbulent pressure and to guide the functional form of the depth-dependence of
the anisotropy of the turbulent velocity field in the 1D stability 
computations.
}
\maketitle
%
\section{Introduction}
\label{intro}
We use LEGACY \cite{LundEtal16} linewidths and frequencies of selected solar-type $Kepler$  stars,
together with 3D hydrodynamical simulation results \cite{TrampedachEtal13, TrampedachEtal14}, 
for calibrating the global stellar and nonlocal convection parameters in 1D stability computations. 
Additionally to exploiting the seismic diagnostic of observed oscillation frequencies,
linewidth measurements provide further diagnostic information about the physical processes
prevailing
in the outer superadiabatic boundary layers where convective transport becomes inefficient.
It is in these outer stellar layers where solar-like oscillations are excited stochastically
to the observed oscillation amplitudes and damped by various processes including the interaction
between oscillations and convection. Therefore, a time-dependent convection model is required for
describing these physical processes, including the pulsationally perturbations to both the convective
heat (enthalpy) and momentum (turbulent pressure) fluxes. A consistent inclusion of turbulent 
pressure in the equilibrium structure is, however, only possible within the framework of a 
nonlocal formulation of 
convection \cite{Gough77b}. Such a convection model is adopted here \cite{Gough77a, Gough77b}. 
Within the generally assumed approximations for constructing a 1D convection model, such as the
Boussinesq approximation \cite{SpiegelVeronis60} (for a recent review see \cite{HoudekDupret15}), 
the turbulent fluxes are consistently estimated in both the equilibrium and nonadiabatic pulsation
calculations. A nonlocal convection formulation typically has additional (nonlocal) parameters
which need calibration. Here we use results from 3D hydrodynamical simulations and linewidth
measurements from $Kepler$ data for calibrating these additional (three) nonlocal
convection parameters. We also include an analytical description for the variation of the 
anisotropy of the turbulent velocity field, the functional form of which being guided by
3D simulations \cite{TrampedachEtal13}. 

\section{Model computations}
\label{computations}
The 1D nonlocal model calculations are carried out 
essentially in the manner described by \cite{HoudekEtal99} and \cite{ChaplinEtal05}
but include, in addition, a description for the variation of the turbulent 
velocity anisotropy with stellar depth, and a temperature-optical-depth ($T-\tau$) relation
derived from 3D simulations \cite{TrampedachEtal14}.
The convective heat flux and turbulent pressure are obtained from a 
nonlocal generalization of the mixing-length formulation \cite{Gough77a, Gough77b}. 
In this generalization three more (nonlocal) parameters, $a, b$ and $c$, are 
introduced which control the spatial coherence of the ensemble of eddies contributing
to the total convective heat flux ($a$) and turbulent pressure ($c$), and the 
degree to which the turbulent fluxes are coupled to the local stratification ($b$).
The effects of varying these nonlocal parameters on the solar structure and oscillation
properties were discussed in detail by \cite{Balmforth92}.

\begin{figure*}[ht]
\centering
\includegraphics[width=\hsize,clip]{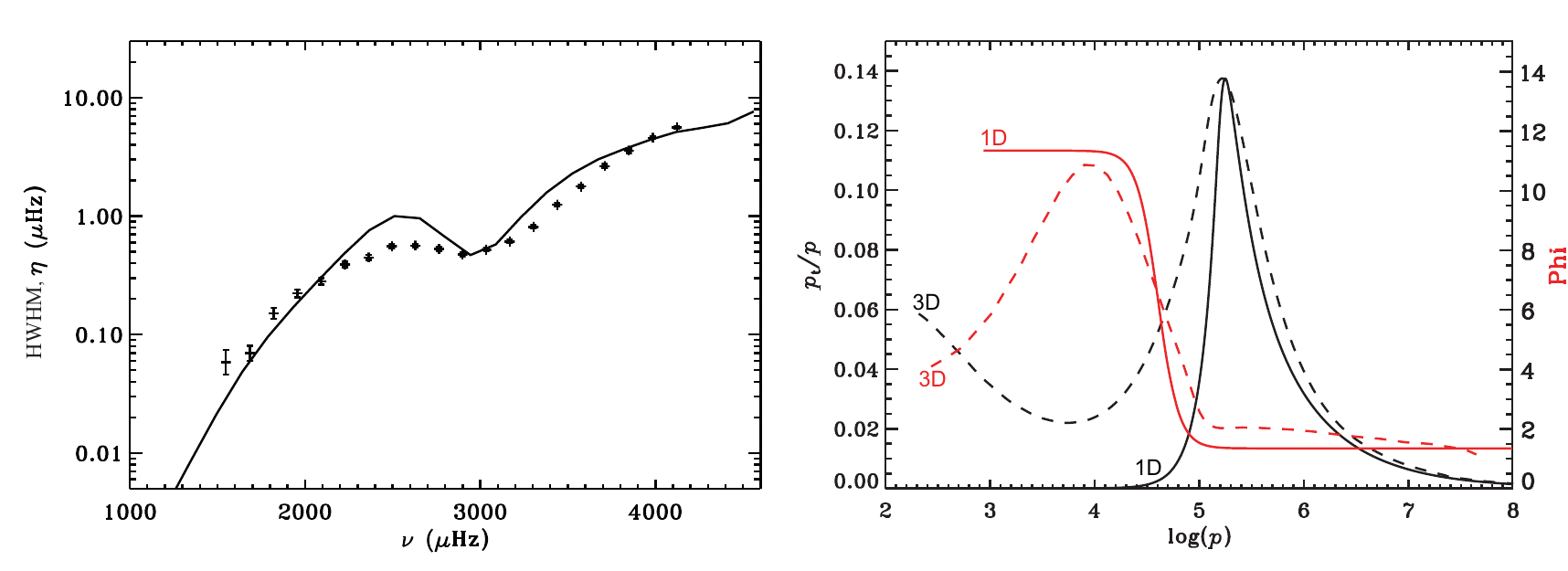}
\caption{Comparison of solar stability computations with BiSON data 
and results from 3D hydrodynamical simulations. The \textbf{left panel} shows theoretical
damping rates $\eta$, in units of cyclic frequency (solid curve), as functions of cyclic 
oscillation frequency $\nu$, together with 
half-width-at-half-maximum (HWHM) measurements (symbols with error bars) of the 
spectral peaks in an acoustic power spectrum obtained by BiSON \cite{ChaplinEtal05}.
The \textbf{right panel} compares profiles of the turbulent pressure $\pt$ over the 
total pressure $p$ (black curves) and turbulent-velocity anisotropy $\Phi$ (red curves)
between a 3D simulation (dashed curves) and a calibrated 1D solar model 
(solid curves). 
}
\label{fig-1}       
\end{figure*}

\begin{figure*}[ht]
\centering
\includegraphics[width=\hsize,clip]{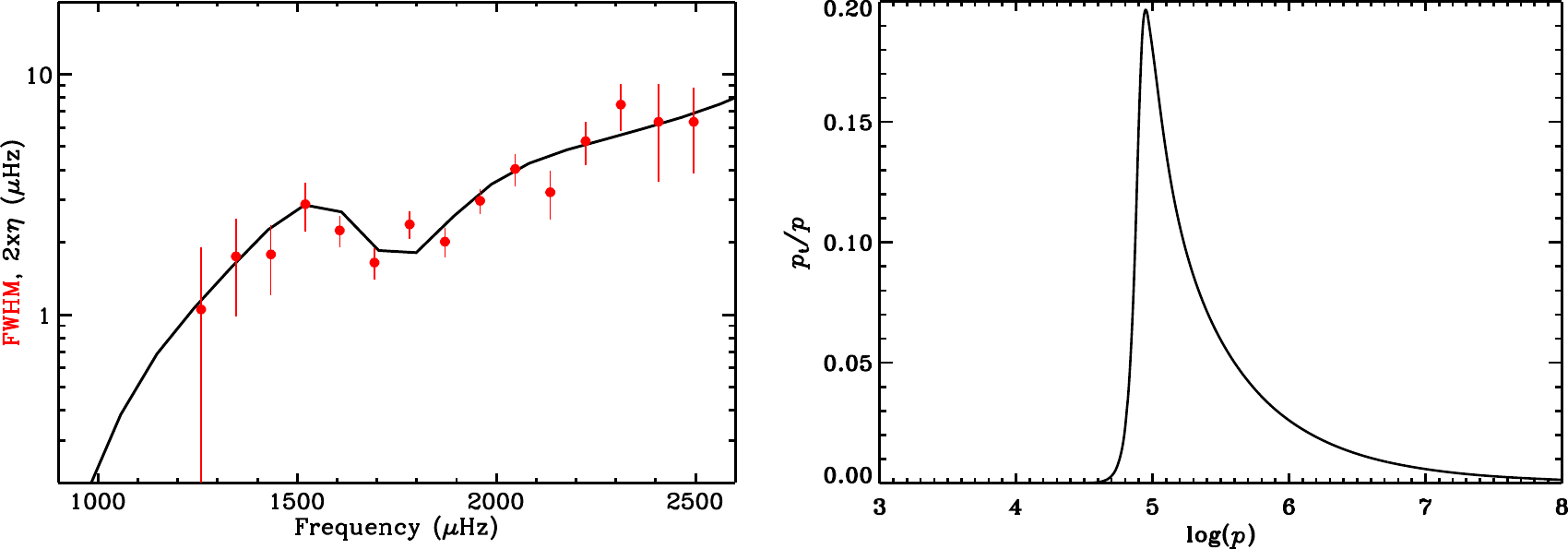}
\caption{Results of a stability calculation for the $Kepler$ star KIC12009504 (Dushera) with 
an effective temperature $T_{\rm eff}$=6217\,K and surface gravity $\log g=$4.214. 
The \textbf{left panel} compares LEGACY linewidths \cite{LundEtal16} 
(red symbols with 1\,$\sigma$ error bars) 
with twice the theoretical damping-rate estimates, 2$\times\eta$ (solid curve), 
in units of cyclic frequency, as functions of cyclic oscillation frequency.
The \textbf{right panel} shows the profile of (3D-calibrated) turbulent pressure $\pt$ over the 
total pressure $p$ in the 1D stellar model.
}
\label{fig-2}       
\end{figure*}

The nonlocal parameter $c$ is calibrated such as to have the
maximum value of the turbulent pressure, max($\pt$), in the 1D nonlocal model 
to agree with the 3D simulation result by \cite{TrampedachEtal13}.
The depth-dependence  of the anisotropy $\Phi:=\overline{\bm{u}\cdot\bm{u}}/\overline{w^2}$ 
of the convective velocity field $\bm{u}=(u,v,w)$ (an overbar denotes an ensemble average) 
is described by an analytical function, guided by 3D simulation results \cite{TrampedachEtal13}. 
The remaining nonlocal parameters $a$ and $b$ cannot be easily obtained from the 
3D simulations and are therefore calibrated such as to have a good agreement between 
calculated damping rates and LEGACY linewidths over the whole measured
frequency range. The mixing length is calibrated such as to obtain the 
surface-convection depth of the frequency-calibrated evolutionary model 
calculated with ASTEC \cite{JCD08a}. 
The ASTEC evolutionary model is 
obtained by minimizing the differences between observed (LEGACY) 
and adiabatically computed (ADIPLS \cite{JCD08b}) oscillation frequencies, including corrections 
for surface effects according to \cite{JCD12}.

\begin{figure*}[ht]
\centering
\includegraphics[width=\hsize,clip]{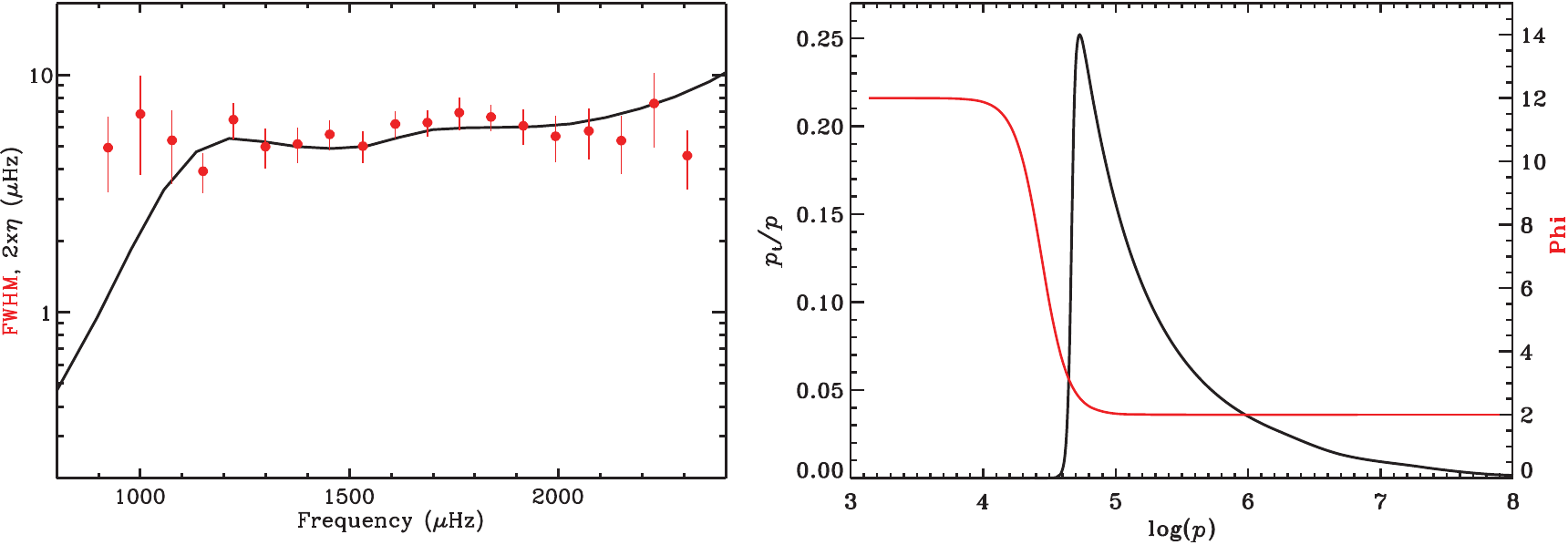}
\caption{Results of a stability calculation for the $Kepler$ star KIC11253226 (Tinky) with 
an effective temperature $T_{\rm eff}$=6715\,K and surface gravity $\log g=$4.171. 
The \textbf{left panel} compares LEGACY linewidths \cite{LundEtal16} 
(full-width-at-half-maximum FWHM, red symbols with 1\,$\sigma$ error bars) 
with twice the theoretical damping-rate estimates, 2$\times\eta$ (solid curve), 
in units of cyclic frequency, as functions of cyclic oscillation frequency.
The \textbf{right panel} shows the profiles of (3D-calibrated) turbulent pressure $\pt$ 
over the total pressure $p$ (black curve) and turbulent-velocity anisotropy $\Phi$ 
(red curve) in the 1D stellar model.
}
\label{fig-3}       
\end{figure*}

\begin{figure}[ht]
\centering
\includegraphics[width=\hsize,clip]{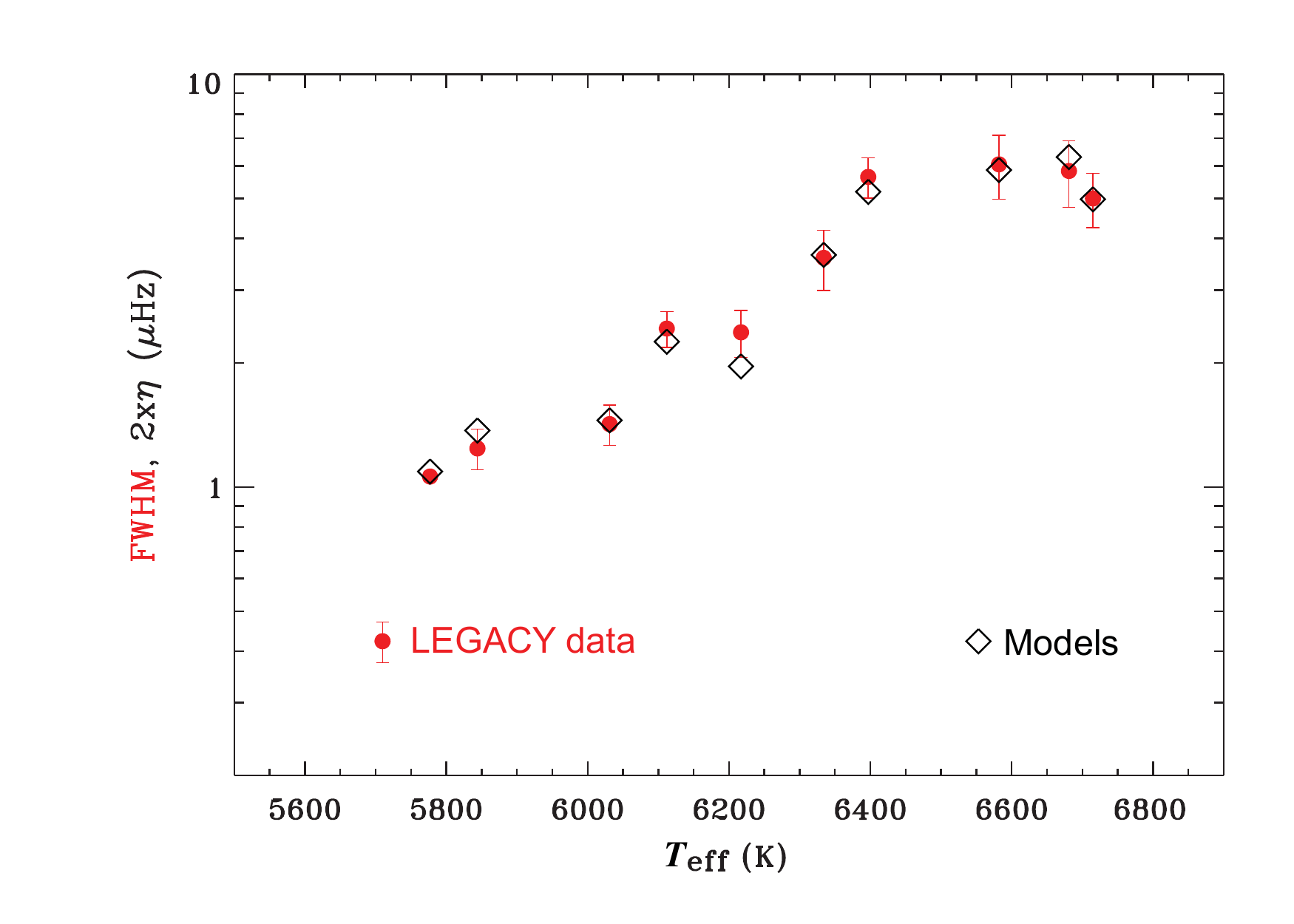}
\caption{Comparison of measured LEGACY linewidths (full-width-at-half-maximum FWHM, 
red, filled circles with 1\,$\sigma$ error bars, \cite{LundEtal16}) 
with twice the estimated damping rates (black, open diamond symbols), in 
units of cyclic frequency, for 9 $Kepler$ stars 
and the Sun (for the Sun BiSON linewidths \cite{ChaplinEtal05} are shown).
Results are plotted as functions of the model's effective temperatures and at 
the frequency $\nu\simeq\nu_{\rm max}\propto gT_{\rm eff}^{-1/2}$, assuming
a solar $\nu_{\rm max}\simeq3100\,\mu$Hz.}
\label{fig-4}       
\end{figure}

Both the nonlocal envelope and pulsation calculations of the stability analyses assume 
the generalized Eddington approximation to radiative transfer \cite{UnnoSpiegel66}.
The temperature gradient in the plane-parallel atmosphere is corrected by 
using a radially varying Eddington factor fitted to $T-\tau$ relations
obtained from 3D simulation results \cite{TrampedachEtal14}.
The abundances by mass of hydrogen and heavy elements are adopted from 
the frequency-calibrated evolutionary calculations.
The opacities are obtained from the OPAL tables
\cite{IglesiasRogers96}, supplemented at low temperature by tables 
from \cite{Kurucz91}.
The equation of state includes a detailed treatment of the ionization
of C, N, and O, and a treatment of the first ionization of the next
seven most abundant elements \cite{JCD82}.
The integration of stellar-structure equations starts 
at an optical depth of $\tau=10^{-4}$ and ends at a radius 
fraction $r/$R$_\odot=0.2$.

The linear nonadiabatic pulsation calculations are carried out using
the same nonlocal convection formulation with the assumption that all eddies in the cascade
respond to the pulsation in phase with the dominant large eddies.
A simple thermal outer boundary condition is adopted at the temperature minimum 
where for the mechanical boundary condition the solutions are 
matched smoothly onto those of a plane-parallel isothermal atmosphere.
At the base of the model envelope the 
conditions of adiabaticity and vanishing of the displacement eigenfunction 
are imposed. Only radial p modes are considered. 

\section{Results}
\label{results}
We first tested the stability analysis against solar data provided by the 
BiSON group \cite{ChaplinEtal05} (see also \cite{HoudekEtal16}). 
The results are illustrated in Figure~\ref{fig-1}. The left panel compares
radial damping-rate estimates $\eta$, in units of cyclic frequency, 
with half-width-at-half-maximum (HWHM) measurements of the spectral peaks 
in the acoustic power spectrum. The power spectrum was obtained from a 
BiSON time-series collected over an epoch of 3456 days \cite{ChaplinEtal05}. 
The right panel compares the profiles of turbulent pressure $\pt$ and 
convective velocity anisotropy $\Phi$ as functions of the logarithm of the 
total pressure $p$ in the outer stellar layers between a 3D solar 
simulation \cite{TrampedachEtal13} and a calibrated 1D model. Only the maximum value of 
$\pt:=\overline{\rho ww}$ ($\rho$ is density) in the 1D model was 
calibrated to the 3D simulation results by adjusting the nonlocal 
convection parameter $c$. The agreement 
between the model and simulation is particularly satisfactorily in the 
deeper stellar layers where the modes are propagating.
Note that in the outer layers with $\log p\lesssim 5.3$ the modes are evanescent
(e.g. \cite{HoudekEtal16}). 
The increasing difference in $p_{\rm t}$ between simulation (dashed curve) and 
model (solid curve) with decreasing total pressure $p$ 
is predominantly a consequence of neglecting in a Boussinesq fluid the acoustic flux 
(H.-G. Ludwig, personal communication), generated by the convective 
fluctuations. These differences in stellar 
stratification have, however, little effect on the acoustic oscillation 
properties, for these differences are confined in the evanescent layers
(see also \cite{HoudekEtal16}).

Next we estimated damping rates in 9 solar-type $Kepler$ stars with effective temperatures
5844\,K\,$\leq T_{\rm eff}\leq6715$\,K. Results for the frequency-dependence of the estimated 
damping rates and profiles of turbulent pressure of the corresponding 1D stellar stratifications
are illustrated in Figures~\ref{fig-2} and \ref{fig-3} for two models with different effective
temperatures. As in the solar case, max($p_{\rm t}$) in the 1D model was calibrated such as 
to match max($p_{\rm t}$) in the 3D simulations \cite{TrampedachEtal13}. 

Figure~\ref{fig-4} compares LEGACY linewidths \cite{LundEtal16} (red, filled circles with error bars)
with estimated damping rates (black, open diamonds) for 9 selected $Kepler$ stars and for the Sun. 
Results are plotted at the frequency $\nu_{\rm max}$ of maximum oscillation amplitude,
determined from the scaling relation $\nu_{\rm max}\propto gT^{-1/2}_{\rm eff}$ 
($g$ being surface gravity) and assuming
a solar $\nu_{\rm max}\simeq3100\,\mu$Hz. The agreement between observations and models
is very satisfactorily. Moreover, all 1D models have global stellar parameters 
determined from minimizing the differences between observed and adiabatically 
computed oscillation frequencies \cite{LundEtal16}, calibrated max($p_{\rm t}$) values, 
depth-dependent functional forms of the turbulent-velocity anisotropy $\Phi$ and 
$T-\tau$ relations as suggested by 3D simulations \cite{TrampedachEtal13, TrampedachEtal14}.  

\section{Acknowledgements}
\label{sec-ack}
We thank Mikkel N. Lund for providing the LEGACY linewidths and 
J{\o}rgen Christensen-Dalsgaard for the global stellar parameters of 
the frequency-calibrated ASTEC evolutionary models.
Funding for the Stellar Astrophysics Centre is provided
by The Danish National Research Foundation (Grant DNRF106).

\begin{thebibliography}{}
%
%

\bibitem[\protect\citeauthoryear{{Balmforth}}{{Balmforth}}{1992}]{Balmforth92}
{Balmforth} N.~J., MNRAS {\textbf{255}, 603} (1992)

\bibitem[\protect\citeauthoryear{{Chaplin}, {Houdek}, {Elsworth}, {Gough},
  {Isaak}  \& {New}}{{Chaplin} et~al.}{2005}]{ChaplinEtal05} 
{Chaplin} W.~J.,  {Houdek} G.,  {Elsworth} Y.,  {Gough} D.~O.,  {Isaak} G.~R.,
   {New} R., MNRAS {\textbf{360}, 859 (2005)}

\bibitem[\protect\citeauthoryear{{Christensen-Dalsgaard}}{{Christensen-Dalsgaard}}{1982}]{JCD82}
{Christensen-Dalsgaard} J., {MNRAS} {\textbf{199}, 735} (1982)

\bibitem{JCD08a}
{Christensen-Dalsgaard} J., Ap\&SS {\textbf{316}, 13} (2008)

\bibitem{JCD08b}
{Christensen-Dalsgaard} J., Ap\&SS {\textbf{316}, 113} (2008)

\bibitem{JCD12}{Christensen-Dalsgaard} J., 
\textit{ESF Conference: ``The Modern Era of Helio- and Asteroseismology''} 
(M. Roth ed., Obergurgel), Astron. Nachr. {\textbf{333}, 914} (2012)

\bibitem[\protect\citeauthoryear{{Gough}}{{Gough}}{1977a}]{Gough77b}
 {Gough} D.~O., \textit{Lecture Notes in Physics} {\textbf{71}},
({Spiegel} E.~A.,  {Zahn} J.-P.,  eds, Springer, New York), 15 (1977a)

\bibitem[\protect\citeauthoryear{{Gough}}{{Gough}}{1977b}]{Gough77a}
{Gough} D.~O., ApJ {\textbf{214}, 196 (1977b)}

\bibitem[\protect\citeauthoryear{{Houdek}, {Balmforth}, {Christensen-Dalsgaard}
   \& {Gough}}{{Houdek} et~al.}{1999}]{HoudekEtal99}
{Houdek} G.,  {Balmforth} N.~J.,  {Christensen-Dalsgaard} J., {Gough} D.~O.,
   AA {\textbf{351}, 582} (1999)

\bibitem[\protect\citeauthoryear{{Houdek} \& {Dupret}}{{Houdek} \&
  {Dupret}}{2015}]{HoudekDupret15}
{Houdek} G.,  {Dupret} M.-A., Living Rev. Sol. Phys., {\textbf{12}:8} (2015)

\bibitem[\protect\citeauthoryear{{Houdek, Trampedach, Aarslev, Christense-Dalsgaard}}
{{Houdek} et~al.}{2017}]{HoudekEtal16}
Houdek G., Trampedach R., Aarslev M.~J., Christensen-Dalsgaard J., 
{MNRAS} {\textbf{464}, L124} (2017)
{\small[\href{http://dx.doi.org/10.1093/mnrasl/slw193}{DOI:10.1093/mnrasl/slw193}]}
 
\bibitem[\protect\citeauthoryear{{Iglesias} \& {Rogers}}{{Iglesias} \&
  {Rogers}}{1996}]{IglesiasRogers96}
{Iglesias} C.~A.,  {Rogers} F.~J., ApJ {\textbf{464}, 943} (1996)

\bibitem[\protect\citeauthoryear{{Kurucz}}{{Kurucz}}{1991}]{Kurucz91}
{Kurucz} R.~L., \textit{NATO Series C} \textbf{341}, 
({Crivellari} L.,  {Hubeny} I.,   {Hummer} D.~G., eds, D. Reidel Publishing Co., Trieste Italy),
 441 (1992)

\bibitem{LundEtal16}
Lund M.~N., Silva Aguirre V., Davies G.~R., Chaplin W.~J., Christensen-Dalsgaard J. et al.,
submitted to ApJ (2016)

\bibitem[\protect\citeauthoryear{{Trampedach}, {Asplund}, {Collet}, {Nordlund}
  \& {Stein}}{{Trampedach} et~al.}{2013}]{TrampedachEtal13}
{Trampedach} R.,  {Asplund} M.,  {Collet} R.,  {Nordlund} {\AA}., {Stein}
  R.~F., ApJ {\textbf{769}, 18 (2013)}

\bibitem[\protect\citeauthoryear{{Trampedach}, {Stein},
  {Christensen-Dalsgaard}, {Nordlund}  \& {Asplund}}{{Trampedach}
  et~al.}{2014a}]{TrampedachEtal14}
{Trampedach} R.,  {Stein} R.~F.,  {Christensen-Dalsgaard} J.,  {Nordlund}
  {\AA}.,   {Asplund} M., {MNRAS} {\textbf{442}, 805 (2014)}

\bibitem[Spiegel and Veronis(1960)]{SpiegelVeronis60}
Spiegel E.~A., Veronis G., {ApJ} {\textbf{131}, 442} (1960)

\bibitem[\protect\citeauthoryear{{Unno} \& {Spiegel}}{{Unno} \&
  {Spiegel}}{1966}]{UnnoSpiegel66}
{Unno} W.,  {Spiegel} E.~A., PASJ {\textbf{18}, 85} (1966)

\end{thebibliography}
%
%

\end{document}